
\def\ie{{\it i.e.}}			
\def\im{{\rm{i}}}			
\def\e{{\rm{e\,}}}			
\def\qb{{{q}_{}}_B}			
\def\qf{{{q}_{}}_F}			

\def\np{Nucl.~Phys.}
\def\pl{Phys.~Lett.}
\def\pr{Phys.~Rev.}
\def\cmp{Commun.~Math.~Phys.}
\def\prs{Proc.~Roy.~Soc.}

\def\M{{{\cal M}_4}}			
\def\MM{{{\cal M}_{4k}}}		
\def\Z{{{\bf Z}}}			
\def\R{{{\bf R}}}                       
\magnification=1200

\advance\hoffset by -3mm       
\advance\voffset by -3mm       

{\line{\hfil{\sevenrm QMW-PH/00-01}}} \vskip -.2cm
{\line{\hfil{\sevenrm SWAT-TH/00-257}}} \vskip -.2cm
{\line{\hfil{\sevenrm ITP-UU-00/05}}} \vskip -.2cm
{\line{\hfil{\sevenrm hep-th/0003155}}}
\vskip .5cm
\centerline{\bf{Spin and abelian electromagnetic duality on four-manifolds}} 
\vskip .5cm
\centerline{David I.~Olive\footnote*{On leave at Instituut voor 
Theoretische Fysica, University of Utrecht, Postbus 80.006, 3508 TA Utrecht}}
\centerline{Department of Physics, University of Wales Swansea,} 
\centerline{Singleton Park, Swansea SA2 8PP, UK}
\centerline{{\tt{e-mail: d.olive@swansea.ac.uk}}}
\smallskip
\centerline{and}
\smallskip
\centerline{Marcos Alvarez}
\centerline{Department of Physics, Queen Mary and Westfield College}
\centerline{Mile End Road, London E1 4NS, UK}
\centerline{{\tt{e-mail: m.alvarez@qmw.ac.uk}}}
\vskip .5cm
\centerline{\bf Abstract}\vskip .1cm
We investigate the electromagnetic duality properties of an abelian
gauge theory on a compact oriented four-manifold by analysing the
behaviour of a generalised partition function under modular
transformations of the dimensionless coupling constants. The true
partition function is invariant under the full modular group but the
generalised partition function exhibits more complicated behaviour
depending on topological properties of the four-manifold concerned.
It is already known that there may be \lq\lq modular weights"
which are linear combinations of the Euler number and Hirzebruch signature
of the four-manifold. But sometimes the partition function transforms
only under a subgroup of the modular group (the Hecke subgroup). In this case
it is impossible to define real spinor wave-functions on the four-manifold.
But complex spinors are possible provided the background
magnetic fluxes are appropriately fractional rather than integral.
This gives rise to a second partition function which enables the full
modular group to be realised by permuting the two partition functions,
together with a third. Thus the full modular group is realised in all cases.
The demonstration makes use of various constructions concerning
integral lattices and theta functions that seem to be of intrinsic interest.


\beginsection 1. Introduction 

It is now widely accepted that electromagnetic duality  
provides a powerful and useful new principle which is valid  
in a large class of physically interesting quantum field theories in a 
Minkowski space-time of four dimensions. The prototype for the quantum version 
of this idea was furnished by a proposal of Montonen 
and one of the present authors, [Montonen and Olive 1977], who considered the 
context of a  special sort of spontaneously broken $SU(2)$ gauge theory, 
later realised to be naturally supersymmetric [D'Adda, Di Vecchia and
Horsley 1978, Witten and Olive 1978, Osborn 1979]. This can be  
regarded as a semi-realistic theory of unified particle interactions 
as it is the same sort of theory as the standard model even  
though it differs in some crucial respects.  The manner in
which duality is realised on particle states requires magnetically
charged states arising as solitons solutions and, in addition,
quantum bound state of these [Sen 1994]. 
  
This acceptance, which now extends to superstring theories, where solitons
occur as higher branes, has been achieved despite the fact that no
sort of proof has been found, even in the case of the $SU(2)$ gauge
theory with the highest allowable degree of supersymmetry, $N=4$. It
is this situation for which most evidence has steadily accumulated
that the idea is exactly true. This supporting evidence has been
facilitated bythe acquisition of new mathematical techniques that have
enhanced our understanding of quantum field theory. For example, the
Atiyah-Singer index theorem, which is related to the theory of axial
anomalies, plays a ubiquitous role. A modifed version of this theorem
is crucial in determining both classical and quantum properties of
self-dual monopole solutions.   
      
Since the quantum electromagnetic duality transformations 
combine to form a group related to the modular group 
(or its generalisations) it is reasonable to expect  
some version of the theory of modular forms to become increasingly 
important. Conversely the idea of electromagnetic duality 
has led to breakthroughs in the classification theory of
four-manifolds (playing the role of Euclidean space-times for twisted
supersymmetric gauge theories).  
 
It would be nice to have a simpler, toy model which, although less
realistic physically,  could, in compensation, be more tractable
mathematically. Indeed, such a theory exists. It is simply free
Maxwell theory with only putative couplings, not realised in
practice. Thus particle states possess what could be called a \lq\lq
Cheshire cat" existence, that will be become clearer later. Such a
theory would be too trivial on flat space time and, in order to obtain
some worthwhile structure, the ambient  space-time manifold has to be
allowed to be fairly general. It is taken to be smooth, compact and
oriented. Hence it is a four-manifold which obeys the topological
symmetry known as Poincar\'e duality, and this will be a crucial
ingredient forestablishing electromagnetic duality here. This toy
model follows the original proposals by E Verlinde and E Witten in 1995.  
 
What is interesting about this model is that much of  
the same sort of mathematical structure as mentioned above  
again comes into play. In particular there enter modular invariant 
theta functions of a rather general nature. The Atiyah-Singer index theorem 
plays a mysterious role interlocked with the modular group 
and its subgroups of index three and the nature of space-time and its possible
spin structures. This is despite the significant differences between
the two situations (no hint of supersymmetry, curved Riemannian
space-time in one case, $N=4$ supersymmetry, flat, non-compact
Minkowski space-time in the other). This points to the conclusion that
electromagnetic duality is indeed a rather general phenomenon. 
 
Moreover it is likely that this picture extends to any space-time with
Minkowski metric and dimension which is a multiple of four. If this
dimension is denoted $4k$ then the putative coupling of Maxwell field
strengths with charged particles is replaced  by a coupling
of a $2k$-form field strength to $(2k-2)$-branes.

In section 2 we start with the naive idea of electromagnetic
duality as a classical symmetry of the energy-momentum tensor
of Maxwell theory in Minkowski space with respect to rotations
between the electric and magnetic fields. It is shown how this idea
can be extended  to a larger group of $SL(2,\R)$ transformations acting 
on the energy momentum tensor. It is explained how this leads 
in turn to consideration of the Feynman path integral, of, rather
surprisingly, the exponential of the Euclidean action. A special case
of this Euclidean path integral is the Minkowski space partition function.

The effect of the Dirac quantisation condition for magnetic fluxes
exhibits some extra subtleties in space-times of four dimensions,
particularly when complex spinor wave functions are considered. This
is reviewed in section 3. The effect is to break the continuous
$SL(2,\R)$ group to a discrete subgroup related to the modular group
in a way that is explained in sections 4 and 5. It is noteworthy that
nothing like the Zwanziger-Schwinger quantisation condition for dyonic
charges plays any role.  

Owing to the complicated topology of the four-manifold of space-time, the
possible magnetic fluxes are related to a  lattice formed by the free
part of the second homology group. This lattice is unimodular by
virtue  of Poincar\'e duality and its even or oddness properties are
related to the presence or absence of spin structures on the space-time
four manifold. In fact four-manifolds can be separated into three
distinct types whose properties are described in section 3. 

In section 4 properties of integral lattices are reviewed. Starting from an odd
unimodular lattice a general construction is given of an even integral
lattice, leading sometimes to new unimodular lattices which can be
both odd or even. Furthermore, also in section 4  the relevance to the Dirac
quantisation condition and the question of spin structures is explained.

In section 5, following the arguments of E Verlinde and E Witten [1995],
the \lq\lq extended partition functions" are evaluated explicitly
using the Dirac quantisation conditions and the semiclassical
method. The results are proportional to generalised theta functions 
associated with the unimodular lattice formed by the free part of
the second homology group. Particular attention is paid to the
different possibilities afforded by the compatibility of either scalar
or spinor complex wave functions on four-manifolds of type II, \ie\
when the relevant lattice is odd. 

In section 6 a more general construction is presented that associates 
theta functions with any integral lattice, not necessarily unimodular,
whether or not the scalar product is positive definite.
Such integral lattices are contained as subgroups of the lattices reciprocal to
them and so define a finite number of cosets, to each of which
corresponds a theta function. An action of the modular group is
defined on these theta functions when the lattice is even and of the
Hecke subgroup when the lattice is odd. Careful analysis of the
self-consistency of this action furnishes a proof of \lq\lq Milgram's formula",
valid for any even integral lattice. This expresses
the signature of the lattice, mod eight, in terms of coset properties.

In section 7, the theta function construction of section 6
is applied to the even integral lattice associated
with an odd unimodular lattice by the construction of section 4.
The result is to associate up to four theta functions with a given
odd unimodular lattice, though there are usually linear dependences.
When the odd unimodular lattice corresponds to that associated with any
type II four-manifold, two of these theta functions enter
the distinct Maxwell partition functions for fluxes
supporting either scalar or spinor complex wave functions, provided
the electric charges they carry coincide. Although each partition function
is individually covariant with respect to only a subgroup of
$SL(2,\Z)$ of index three, they are related to each other by the missing
transformations. In this way the full $SL(2,\Z)$ group of
electromagnetic duality transformations is restored for space-time
four-manifolds of type II. 

\beginsection 2.  Abelian gauge fields and electromagnetic duality

We shall usually be considering a single abelian Maxwell field
strength described, in exterior calculus notation, by a closed two form $F$
on a space-time manifold $\M$ which is closed, compact, connected,
smooth and oriented. Nevertheless much of the argument extends to
higher dimensional space-times of the same type, (which will be
denoted $\MM$), as long as their dimension is a multiple of four and
the field-strength $F$ is a closed $2k$-form, that is a mid-form. 
It would then be a generalised Kalb-Ramond field, [Kalb and Ramond 1974],
that could couple to the world-volume of a $2k-2$-brane.

For reasons that will become clear it is important to allow the
space-time manifold to be topologically complicated. Sometimes we
shall suppose that $\MM$ be such that it can be endowed with a
Minkowski metric (that is with one time component) and this would
require that its Euler number $\chi(\MM)$ vanish. We may further
require that this metric can always be \lq\lq Wick rotated" to a
Euclidean metric (with no time components) by an analytic
continuation. This may impose further constraints on the topological
properties of $\MM$. 

With either sort of metric the Hodge star  operation, $*$, can be
defined, converting $p$-forms on $\MM$ to $4k-p$-forms. Acting on
$2k$-forms, that is, mid-forms, the repeated action of the Hodge star
operation yields 
$$
**=(-1)^t,
\eqno(2.1)
$$ 
where $t$ equals the number of time components, that is, one or
zero. So, in the Minkowski case, $*$ has eigenvalues $\pm \im$ and
hence no real eigenfunctions. It is this Minkowski situation, rather
than the Euclidean one, in which electromagnetic duality can apply. 

The basic idea of a resemblance between the parts of $F$ thought of as
the electric and magnetic fields, $\underline E$ and $\underline B$,
is very old and was reinforced by Maxwell's discovery of his equations
governing their behaviour in vacuo. \lq\lq Duality rotations" between
$\underline E$ and $\underline B$ provide a symmetry of the Minkowski
energy density $(E^2+B^2)/2$ (this particular expression applies when
the space-time is flat), and more generally the complete energy
momentum tensor. Moreover the rotations map between solutions of the
equations even though the action does change. 

This idea was extended in the context of supergravity theories 
by Gaillard and Zumino [1981], building on ideas of Cremmer and Julia [1979].
The following action can be defined on any of the space-times mentioned:
$$
W={1\over2\tau_2}\int_{\M}F\wedge \hat\tau F,
\eqno(2.2)
$$
where, in Minkowski space,
$$
\hat\tau=\tau_1+*\tau_2={\theta\over2\pi}+*{2\pi\hbar\over q^2},
\eqno(2.3)
$$
so that this action, (2.2), is indeed real. In suitably extended
supergravity theories $\tau$ will depend upon scalar fields related to
the metric tensor by supersymmetry transformations. As far as this
paper is concerned, there are no scalar fields and no
supersymmetry. $\tau_1$ and $\tau_2$ are simply dimensionless
quantities parametrising the theory. It will be convenient to combine them
as real and imaginary parts of a single complex variable
$$
\tau=\tau_1+i\tau_2={\theta\over2\pi}+\im\,{2\pi\hbar\over q^2}
\eqno(2.4)
$$ 
The conventional Maxwell term is ${1\over2}\int F\wedge*F$. The other
term, called the theta term, has no apparent effect classically as it
affects neither the Euler-Lagrange equations, nor the value of the
energy-momentum tensor. However it can affect the quantum phase when
the topology of the background space-time is sufficiently
non-trivial. In circumstances to be explained,
$(\tau_2)^{-1}\int_{\M}F\wedge F$ is quantised so that the
exponentiated quantum action, $\exp({\im\over\hbar}W)$, depends upon the
parameter $\theta$ in a periodic manner. But the dependence upon
$\theta$ disappears altogether when the topology is too trivial, 
that is, when the second Betti number of $\M$ vanishes.

Following Gaillard and Zumino we consider the effect of the linear
transformations: 
$$
\hat\tau F\rightarrow \hat\tau ' F'=A\hat\tau F+BF
\eqno(2.5a)
$$
$$
F\rightarrow F'=C\hat\tau F+D F,
\eqno(2.5b)
$$
where $A, B, C$ and $D$ are real constants. 

Then the dimensionless complex coupling constant variable, $\tau$, (2.4)
undergoes the fractional linear transformation
$$
\tau\rightarrow \tau'={A\tau+B\over C\tau+D}.
\eqno(2.6a)
$$
while its imaginary part $\tau_2$ undergoes
$$
\tau_2\rightarrow \tau_2'={\tau_2(AD-BC)\over|C\tau+D|^2}.
\eqno(2.6b)
$$
The transformations (2.5) evidently provide symmetries of the two
equations of motion $dF=0$ and $d\,\hat{\tau}F=0$. Now we consider the
effect on the symmetric energy momentum tensor $T_{\mu\nu}$, obtained
from (2.2) by variation of the metric and written in the Sugawara form,
$$
T_{\mu\nu}={1\over2}(F_{\mu\lambda}\,g^{\lambda\sigma}\,F_{\sigma\nu}+
^*F_{\mu\lambda}\,g^{\lambda\sigma}\,^*F_{\sigma\nu}),
\eqno(2.7)
$$
where $F=F_{\mu\nu}\,dx^{\mu}\wedge dx^{\nu}/2$ and
$^*F=^*F_{\mu\nu}\,dx^{\mu}\wedge dx^{\nu}/2$. The result is
$$
T_{\mu\nu}\rightarrow T_{\mu\nu}'=|C\tau+D|^2T_{\mu\nu}.
$$
If we restrict the transformations (2.5) to the subgroup leaving
$\tau$ unchanged in (2.6), evidently $\tau_2$ is also unchanged and so
by (2.6b), the result is 
$$
T_{\mu\nu}\rightarrow T_{\mu\nu}'=(AD-BC)T_{\mu\nu}.
$$
Hence the energy momentum tensor is invariant under the subgroup if 
$$
AD-BC=1
\eqno(2.8)
$$
and this yields a $U(1)$ subgroup comprising the duality rotations 
previously mentioned. The symmetry of the energy momentum tensor (2.7)
can be enlarged from this $U(1)$ to the full $SL(2,\R)$ if the
transformations (2.5) are modified in the following natural way. Let
us substitute for the physical field strengths $F$ more geometrical
quantities $G$, by $F=\hbar G/q$. $G$ is more geometrical in the
sense that its fluxes, unlike those of $F$ are dimensionless. In terms
of $G$ the dimensionless action is given by 
$$
{W\over\hbar}={1\over4\pi}\int_{\M}G\wedge\hat{\tau}G
$$
and so is explicitly a function of $\tau$. Now the energy momentum
tensor (2.7) reads 
$$
T_{\mu\nu}={\hbar\tau_2\over4\pi}(G_{\mu\lambda}\,g^{\lambda\sigma}\,
G_{\sigma\nu}+ ^*G_{\mu\lambda}\,g^{\lambda\sigma}\,^*G_{\sigma\nu}).
\eqno(2.7')
$$
so that a dependence on $\tau_2$ is made explicit.
Now substitute $G$ for $F$ in (2.5), thereby defining new duality
transformations, still leading to (2.6). Under these transformations
acting on both $G$ and $\tau$ in the energy momentum tensor $T_{\mu\nu}$
in the form (2.7'), the preceding calculations show that $T_{\mu\nu}$,
is invariant providing only that (2.8) holds. 

Thus now the dimensionless complex coupling $\tau$ changes
whilst preserving the positive nature of $\tau_2$. This is appropriate
as $\tau_2$ is the inverse of the fine structure constant and hence
intrinsically positive.  

These transformations form the three dimensional non compact group
$SL(2,\R)$, or what is the same by a group theory isomorphism, the
symplectic group $Sp(2,\R)$. 

The transformations  also map between solutions of the free Maxwell
equations and the key question will be to what extent these
transformations continue to provide symmetries when there is a possibility of
electrically charged particles (or branes) being present, subject to
the rules of quantum theory. Because it is the Minkowski energy,
rather than the action, which is invariant classically, the natural
quantity to consider in the quantum theory is the partition function
constructed with this energy: 
$$
Z(\tau)=Tr\left(\e^{-E(\tau)}\right).
\eqno(2.9)
$$
It is this partition function that will be the candidate for quantum 
electromagnetic duality, just as it is the partition function
that displays the Kramers-Wannier duality of the Ising model
[Kramers and Wannier 1941]. Indeed it will be found that (2.9) is
invariant under the transformations (2.6) provided they are restricted
to a discrete subgroup, isomorphic to the modular group. The
discreteness is a consequence of the Dirac quantisation condition that
the magnetic fluxes have to satisfy in order to permit complex wave functions. 

For rather general, nonlinear dynamical systems with a finite number
of degrees of freedom and the property that the action includes only
terms quadratic, linear and independent of velocities there is a
Feynman path integral expression for the partition function (2.9).
$$
Z(\tau)=\int\dots\int\delta A\,\e^{{\im\over\hbar}W_{EUCLIDEAN}}.
\eqno(2.10)
$$
The Euclidean action $W_{EUCLIDEAN}$ is obtained from the original action 
by what can be thought of as a \lq\lq Wick rotation" whereby velocities are
multiplied by $\im$ and time by $-\im$. As a result, $iW_{EUCLIDEAN}$ has
an imaginary part linear in velocities, and a real part that is negative
definite if the original energy is positive. Consequently the path integral
is highly convergent. Because of the trace in (2.9), the paths integrated over
are closed paths traversed in configuration space in unit time with
distinguished end points. This result is known as the Feynman-Kac
formula [Feynman and Hibbs, Feynman].

The presence of a complex phase factor in (2.10) due to terms linear
in velocity appears to contradict
the manifest reality of the partition function as defined in
(2.9). But this is illusory because the space of closed paths in
configuration space that are integrated over possess a $Z_2$ symmetry
with respect to the interchange of pairs of identical paths
differing only in the sense of time evolution along the path.
Under this interchange the two contributions to
$\exp{{\im\over\hbar}W_{EUCLIDEAN}}$ are related by complex
conjugation. As a result the sum of these two complex contributions is
indeed real. 

Because it is quadratic in field strengths, something similar happens
with the more complicated action (2.2) under consideration here. As a
result of a similar argument, the Maxwell partition function can be
expressed in the form (2.10) where now $W_{EUCLIDEAN}$ is obtained
from $W$, (2.2), by a \lq\lq Wick rotation" of the metric tensor. This
tensor only enters the Maxwell term as the theta term is \lq\lq
topological" and hence independent of the metric, and so unaffected by
the Wick rotation. The result is that $W_{EUCLIDEAN}$ is given by the
same expression as before, (2.2), when it is understood that $\tau$ is
replaced by 
$$
\tau_{EUCLIDEAN}=\tau+\im *\tau_2=
{\theta\over2\pi}+\im *{2\pi\hbar\over q^2}.
\eqno(2.11)
$$
The metric dependence is encoded in the Hodge $*$ operator which, by (2.1),
now has unit square and hence eigenvalues $\pm1$. This has two consequences. 
One is that $\im W_{EUCLIDEAN}$ is complex when the field strengths are real 
and  that its real part is negative definite, thereby ensuring convergence of
the integral over gauge potentials $A$ in (2.9). The other consequence
is that if $*$ is regarded as imaginary in Minkowski space-time and
real in Euclidean space, in view of its eigenvalues, then $\tau$ has
the same complex structure in either case. In this sense it is
unaffected by the Wick rotation. Accordingly the complex variable
$\tau$ given by the expression (2.4) not involving the Hodge $*$ is
equally relevant with either metric. 

In evaluating this partition function the space-time four manifold
has to be considered as $\M=S_1\times{\cal M}_3$, where \lq\lq time"
is the coordinate around the circle, (periodic because of the trace),
and ${\cal M}_3$ an appropriate section of $\M$. If the metric on
$S_1\times{\cal M}_3$ factorises correspondingly it is easy to see
that the partition function will be again real because reversing the
sense of time around this circle will effect a complex conjugation of the two
quantum amplitude contributions.

It will turn out to be highly instructive to consider what, by abuse
of terminology, is often also called a \lq\lq partition function".
This expression is given  by the path integral (2.10) but with the
integral in the action being over the four manifold given by
the full space-time $\M$, instead of $S_1\times{\cal M}_3$. This path
integral can be defined even for four manifolds with non-vanishing
Euler number, that is ones for which a Minkowski metric is impossible.
There is no reason for this new quantity to be real but it will turn
out to have an interesting response to the electromagnetic duality
transformations (2.6) (as pointed out by E Witten and E Verlinde). So
these extended partition functions do have interesting mathematical
properties as we shall see in more detail and it would be interesting
to understand what, if any, physical significance they have. We shall
henceforth refer to the real partition functions associated with
$S_1\times{\cal M}_3$ as \lq\lq strict partition functions". 

We shall see that both the Euler number, $\chi(\M)$, and the
Hirzebruch signature, $\eta(\M)$, vanish for manifolds $S_1\times{\cal
M}_3$. These are the two topological invariants of a $\MM$ manifold
that are \lq\lq local" in the sense that they can be expressed as
integrals of closed forms over the manifold. Linear combinations of
these topological invariants, namely $(\chi\pm \eta)/2$, will specify
in a precise way how the extended partition functions deviate from
satisfying exact electromagnetic duality. 

These conclusions will depend on the explicit evaluation of the
functional integral expression (2.10) for the partition function on
any $\M$, and this is facilitated by taking account of another aspect
of the quantum theory. If electrically charged particles are
to be treated quantum mechanically, the background field strengths
must satisfy certain Dirac flux quantisation conditions in
order to allow the possible presence of complex  wave functions for
them. It is this that imposes a discrete structure that converts (2.10) to
a sum rather than an integral, at least in the semiclassical
approximation, which is very likely exact.

As we shall see, the resultant expression for the partition function
(2.9) is proportional to an infinite sum forming a generalised sort of
theta function associated with the lattice of homology classes of two-cycles
in the space-time four-manifold $\M$. As explained below, this lattice is
what is known as the free part of $H_2(\M,\Z)$ and is intimately connected to
the Dirac quantised fluxes. 

If $\M$ is orientable, smooth, closed and compact, its topological structure
satisfies the symmetry known as Poincar\'e duality. This implies that
the following relation between the five Betti numbers of $\M$:
$$
b_0=b_4,\qquad b_1=b_3.
\eqno(2.12)
$$ 
Hence the Euler number is given by
$$
\chi(\M)=2(b_0-b_1)+b_2.
\eqno(2.13)
$$
Furthermore, the aforementioned lattice, which has dimension $b_2$, 
is unimodular with respect to the scalar product furnished by the
intersection number. It is this which is the origin of the covariance of 
(2.9) with respect to the $S$-transformation of electromagnetic duality:
$$
S=\pmatrix{0&-1\cr1&0\cr}
\eqno(2.14)
$$
sending $\tau$ to $-1/\tau$, a especially interesting example of
(2.6). Before explaining this we must review the Dirac quantisation
condition for fluxes in more detail, paying particular attention to
the extra subtleties associated with spinning particles carrying
electric charge.

\beginsection 3. The quantisation condition on four-manifolds

As already explained, we consider Maxwell theory in space-times that are
compact, connected and oriented four-dimensional manifolds $\M$,
whether or not their Euler number vanishes. These spaces
are particularly convenient because they satisfy Poincar\'e duality,
which is a topological property closely related to electromagnetic
duality. The main mathematical tool needed in understanding the
implications of the global topology of $\M$ on the duality properties
of Maxwell theory is homology and cohomology theory. This discipline
is described in many textbooks (see for example [Schwarz 1994]), and a
short introduction to the relevant ideas was given in our earlier
paper [M Alvarez and Olive 1999]. We shall follow the notations of the
latter without full explanation.

The physical relevance of homology and cohomology theory is that it
provides the natural mathematical language for the ideas of Faraday,
Maxwell and, later, Dirac concerning electromagnetic theory. Important
physical quantities are the (magnetic) fluxes of the field strength
$F$ through a complete set of two-cycles $\Sigma_1, \dots\Sigma_{b_2}$ within 
the space-time four manifold $\M$. 

According to Poincar\'e's lemma, the gauge potential $A$, satisfying
$F=dA$, can be constructed by integration locally, in topologically trivial
neighbourhoods of $\M$, up to a gauge transformation. $A$ is needed to
define the electromagnetic coupling to complex wave-functions of
electrically charged particles. If the wave function is scalar,
corresponding to a boson with charge $\qb$, quantum mechanical
consistency of the patching procedure for it requires the fluxes to be
quantised [Dirac 1931, Wu and Yang 1975, O Alvarez 1985]:
$$
{\qb\over2\pi\hbar}\int_{\Sigma}F=m(\Sigma)\in\Z
\eqno(3.1)
$$
By virtue of ordinary Stokes' theorem and that the facts that $\Sigma$ and $F$ 
are both closed, the value of the flux is unchanged either if $\Sigma$
is replaced by $\Sigma+\partial\Pi$, with $\Pi$ a three-dimensional
chain, or if $F$ is replaced by $F+dB$. Also, integer linear
combination of cycles that satisfy the quantisation condition $\Sigma$
also satisfy the same condition. 

These statements can be summarised by saying that the cycles $\Sigma$
are free elements of the integer homology class $H_2(\M,\Z)$, and $F$
is in a cohomology class $H^2(\M,\Z)$. Now $H_2(\M,\Z)$ is an abelian
group (with respect to the natural addition operation) and it
possesses a unique subgroup built of elements of finite order, called
the torsion group $T_2(\M,\Z)$. The quotient group
$$
F_2(\M,\Z)\equiv H_2(\M,\Z)/T_2(\M,\Z)
\eqno(3.2)
$$
is \lq\lq free" and consists of $b_2$ copies of the integers, where $b_2$ is 
the second Betti number. This is the same as saying that $F_2(\M,\Z)$ is
a lattice of dimension $b_2$.

A slightly more general version of (3.1) is the \lq\lq quantum Stokes' 
relation":
$$
\e^{\im{q_B\over\hbar}\int_{\Sigma}F}=
\e^{\im{q_B\over\hbar}\int_{\partial\Sigma}A}.
\eqno(3.3)
$$
Now $\Sigma$ is allowed to have a non-vanishing boundary
$\partial\Sigma$ and hence be a two-chain rather than a two-cycle. In
the limiting case when the boundary vanishes, (3.1) is recovered (and
so is necessary for the validity of (3.3)). The quantity on the right
hand side of (3.3) is Dirac's path dependent phase factor [Dirac 1955].
 It is well defined
in the situation described even though the exponent is not. Such phase
factors are relevant in several different contexts [Bohm-Aharonov,
Wilson] and go by several other names (Wilson loop, $U(1)$ holonomy etc).

However, many electrically charged particles, such as the electron,
also carry spin and, as a result (3.1) and (3.3) may have to be
modified if the topology of space-time is sufficiently complicated. If
the complex wave function to which $A$ couples is spinor rather than
scalar, and the associated fermionic particle carries charge $\qf$, (3.3)
is modified by the presence of a possible minus sign [M Alvarez and
Olive 1999]. 
$$
\e^{\im{q_F\over\hbar}\int_{\Sigma}F}=(-1)^{w(\Sigma)}\,\,
\e^{\im{q_F\over\hbar}\int_{\partial\Sigma}A}, 
\eqno(3.4)
$$
at least  for two-chains $\Sigma$ whose boundary is an even cycle:
$$
\partial\Sigma=2\alpha.
\eqno(3.5)
$$
There are essentially two possibilities for this when $\Sigma$ is
odd. Either $\alpha$ vanishes and $\Sigma$ is a closed surface, or
not. If the latter, $\Sigma$ could be the real projective plane in two
dimensions, and so not orientable. Notice that although $2\alpha$ is
closed, $\alpha$ itself is not. Hence the one-cycle $\alpha$ is what
is known as a torsion cycle.

The sign factor $(-1)^{w(\Sigma)}$ in (3.4) arises unambiguously in
the procedure of patching together the neighbourhoods that make up
$\Sigma$, precisely when $\Sigma$ satisfies (3.5), [M Alvarez and Olive 1999].
However, when $\Sigma$ is closed, $w(\Sigma)$ can be constructed
independently as the integer specifying the self-intersection number
of $\Sigma$ with itself. This is possible because $\Sigma$ is a
two-cycle in a closed oriented four-manifold. The equivalence of the
two notions (mod $2$) can be deduced from the Atiyah-Singer index
theorem on $\M$. When $\Sigma$ is neither closed nor even, yet satisfies
(3.5), it is not oriented and its self intersection number can only be
defined mod $2$, and not absolutely. This still matches the previous
definition, according to Wu's formula. The important point is that the
sign factor depends only on the topology of the background space-time.

Let us temporarily set the charges $\qb$ and $\qf$ equal. When the
sign factor $(-1)^{w(\Sigma)}$ equals $1$ equations (3.3) and (3.4)
agree but when it equals $-1$ they appear to differ. However the discrepancy
is illusory as the gauge potentials in each equation differ as they
are constructed by gauge inequivalent patching procedures. 

On the other hand, when we consider the limiting case in which $\qb$
and $\qf$ both vanish the two versions of \lq\lq quantum Stokes'", 
(3.3) and (3.4), reduce to $1=1$ and $1=(-1)^{ w(\Sigma)}$,
respectively. Now the second equation is manifestly a contradiction if
the sign factor is negative. What this means is that a false
assumption has been adopted, namely that it is possible to place an
electrically neutral spinor wave function on $\M$. Clearly this is forbidden
if there is any  $\Sigma$ satisfying (3.5) for which the sign factor
$(-1)^{ w(\Sigma)}$ is negative. Mathematicians are familiar with this 
phenomenon and say that such an $\M$ lacks a \lq\lq spin
structure". They recognise $w(\Sigma)$, as the Stiefel-Whitney
class. It is an element of $H^2(\M,\Z_2)$ and its nontriviality
provides an obstruction to spin structures. See Appendix A of
[Lawson and Michelsohn].

In our previous paper we found it convenient to separate
all the four-manifolds under consideration into three types: I, II and III.

A four-manifold is of Type I if the sign factor $(-1)^{w(\Sigma)}$
is plus one in all cases (3.5). The the flux quantisation condition reads
$$
{\qf\over 2\pi\hbar}\int_{\Sigma}F=m(\Sigma)\in\Z.
\eqno(3.6)
$$
This is the same as (3.1) with $\qf$ replacing $\qb$. It follows that
all cycles have even self-intersection number. 

The intersection numbers of pairs of two-cycles endow the lattice
$F_2(\M,\Z)$ of free cycles with an integral scalar product. The
Poincar\'e duality that applies to the four-manifolds under
consideration implies that this scalar product is
unimodular. Unimodular lattices fall naturally into two classes even,
or odd. It follows from our remarks that the unimodular lattice
$F_2(\M,\Z)$ is even for Type I manifolds.

A four-manifold is of Type II if the sign factor  $(-1)^{w(\Sigma)}$
is minus one for at least one two-cycle $\Sigma$. Such a cycle has an 
odd self-intersection number and consequently the unimodular lattice
$F_2(\M,\Z)$ is odd.

The flux quantisation implied by (3.4) reads
$$
{\qf\over 2\pi\hbar}\int_{\Sigma}F=
m(\Sigma)+{w(\Sigma)\over2}; \qquad m(\Sigma)\in\Z.
\eqno(3.7)
$$
Thus, when $w(\Sigma)$ is odd, the flux is fractional rather than integral
and, in particular, can never vanish, unlike the integral fluxes.

The remaining possibility, Type III, arises when the sign factor  
$(-1)^{w(\Sigma)}$ equals plus one for all cycles but minus for at
least one of the open two-chains satisfying (3.5). As the
self-intersection numbers of all cycles are even so is the unimodular
lattice $F_2(\M,\Z)$. 

Thus, in summary, type I manifolds are the only ones that support spin
structures, that is electrically neutral spinor wave functions. All
three types support \lq\lq spin$_C$" structures, that is electrically
charged complex spinor wave functions, provided the backgound fluxes 
satisfy the appropriate quantisation conditions.

These conditions imply that all fluxes are integral if the unimodular
lattice $F_2(\M,\Z)$ is even, that is for types I and III, but that
some fluxes at least must be fractional when the lattice is odd,
that is for Type II manifolds. Standard examples of the three types of
four-manifold are, for type I the torus, $T_4$, the sphere $S_4$ and
$K(3)$, for type II, the complex projective space, $CP(2)$ and for
type III, $S_2\times S_2/Z_2$.

\beginsection 4. Integral lattices

Unimodular lattices are a  sort of integral lattice with special
structural features that will play a role in the picture of flux
quantisation that we have begun to explain. Here we pause to explain
some of the relevant concepts. 

A lattice $\Lambda$ of dimension $n$ is a discrete subgroup of $\R^n$ 
defined as
$$
\Lambda=\left\{ \sum_{i=1}^n n_i\,e_i,\quad n_i\in\Z\right\},
\eqno(4.1)
$$
where $e_1,e_2,\dots, e_n$ are elements  of $\Lambda$ spanning $\R^n$
and so providing a basis for $\Lambda$.

The vector space $\R^n$ may be endowed with a real, symmetric scalar product, 
denoted $x\cdot y$, which is nonsingular but not necessarily positive
definite. If this scalar product has $b^{\pm}$ positive (negative) eigenvalues
the signature $\eta$ of $\Lambda$ is
$$
\eta(\Lambda)=b^+-b^-, \qquad \hbox{where} \quad b^++b^-=n.
\eqno(4.2)
$$
For example, when $F_2(\M,\Z)$ is endowed with the scalar product
given by the intersection numbers, its signature is known as the Hirzebruch
signature of $\M$, and denoted $\eta(\M)$.

Given a lattice and a scalar product we can define another lattice,
known as the reciprocal lattice:
$$
\Lambda^*=\left\{x\in\R^n: y\cdot
x\in\Z\,\,\forall\,y\in\Lambda\right\}.
\eqno(4.3)
$$
Obviously $\Lambda^{**}=\Lambda$. $\Lambda$ is said to be integral if
$$
\Lambda \subseteq \Lambda^*
\eqno(4.4)
$$
as a subgroup. Then the quotient group is well-defined and abelian
$$
Z(\Lambda)=\Lambda^*/\Lambda.
\eqno(4.5)
$$
$\Lambda$ is unimodular if its order, $|Z(\Lambda)|$, is one. This
is the same as saying that the lattices $\Lambda$ and $\Lambda^*$ coincide.

A convenient choice of fundamental domain in $\Lambda$ consists of 
$X=\sum_i x_i\,e_i$ with $0\leq x_i\leq 1$. The volume of this
fundamental domain is 
$$
V(\Lambda)=\sqrt{\det(e_i\cdot e_j)}=|\det(e_1,e_2,\dots, e_n)|.
$$ 
A standard basis for the reciprocal lattice is the reciprocal basis
$f_1,f_2\dots, f_n$ satisfying $e_i\cdot f_j=\delta_{ij}$. Then
$$
V(\Lambda^*)V(\Lambda)=
|\det(f_1,f_2,\dots, f_n)\det(e_1,e_2,\dots, e_n)|=|\det(e_i\cdot f_j)|=
\det(\delta_{ij})=1.
$$
But $|Z(\Lambda)|$ copies of the fundamental domain of $\Lambda^*$ make
up a fundamental domain for $\Lambda$. Hence
$$
V(\Lambda)=(V(\Lambda^*))^{-1}=\sqrt{|Z(\Lambda)|}.
\eqno(4.6)
$$
Thus $\Lambda$ is unimodular if it is integral and the volume of its
fundamental domain equals unity. This is what was used above. 

In fact any choice of fundamental domain has the same volume. So
changes of basis form the infinite discrete group $SL(n,\Z)$.

If $x$ is an element of an integral lattice $x\cdot x$ is
automatically an integer. If it is always an even integer $\Lambda$
is said to be even. Otherwise $\Lambda$  is odd.

Of course this applies to unimodular lattices but these possess an
extra feature, the existence of elements called characteristic
elements. $c\in\Lambda$ is a characteristic vector if
$$ 
c\cdot x+x\cdot x\in 2\Z\qquad \forall\, x\in\Lambda.
\eqno(4.7)
$$
It is easy to establish that such quantities always exist and that
there is an ambiguity of precisely mod $2\Lambda$. It follows that
$c\cdot c$ is uniquely defined, mod $8$ for any unimodular lattice. In
fact it is known that 
$$
c\cdot c=\eta(\Lambda)+8\Z
\eqno(4.8)
$$
This will follow from our discussion of theta functions but a direct
proof can be found in [Milnor and Husemoller]. The zero element is
always a characteristic vector for an even unimodular lattice. Hence
(4.8) implies that for such lattices the signature is a multiple of
eight. A famous example is the $E_8$ root lattice. 

The fact that an odd unimodular lattice $\Lambda$ possesses
a characteristic vector $c$ will facilitate some constructions which
will be relevant to the analysis of type II four manifolds. First:
$$
\Lambda_{TOTAL}=\Lambda\cup(\Lambda+c/2)
\eqno(4.9)
$$
defines a lattice. Obviously $\Lambda_{TOTAL}/\Lambda=Z_2$, so that, by a
slight extension of (4.6), 
$V(\Lambda_{TOTAL})=1/2$. Furthermore we can split
$$
\Lambda =\Lambda_{EVEN}\cup\Lambda_{ODD}
\eqno(4.10)
$$
where $\Lambda_{EVEN/ODD}$ consists of those elements of $\Lambda$
with even/odd squared length. Now $\Lambda/\Lambda_{EVEN}=Z_2$ and so
$V(\Lambda_{EVEN})=2$. Thus $c$ itself will be in $V(\Lambda_{EVEN/ODD})$
according as $c^2$ is even or odd, or according to (4.8), as the signature
(4.2) is even or odd. It is easy to see that $\Lambda_{EVEN}$ and
$\Lambda_{TOTAL}$ are a pair of reciprocal lattices. Furthermore we have 
$$ 
Z(\Lambda_{EVEN})=\Lambda_{TOTAL}/\Lambda_{EVEN}=\cases{Z_4& if $  
c\in\Lambda_{ODD}$;\cr
Z_2\times Z_2& if $ c\in \Lambda_{EVEN}$.\cr}
\eqno(4.11)
$$
This follows because the quotient group is determined by the addition rules
for the  relevant cosets in the decomposition
$$
\Lambda_{TOTAL}=\Lambda_{EVEN}\cup(\Lambda_{EVEN}+c/2)\cup\Lambda_{ODD}
\cup(\Lambda_{ODD}+c/2)
\eqno(4.12)
$$
It follows from (4.11) that, when $c\in\Lambda_{EVEN}$,
$$
\Lambda'\equiv\Lambda_{EVEN}\cup(\Lambda_{ODD}+c/2),
\quad\Lambda''\equiv\Lambda_{EVEN}\cup(\Lambda_{EVEN}+c/2)
\eqno(4.13)
$$
are both lattices and that $V(\Lambda')=V(\Lambda'')=V(\Lambda_{EVEN})/2=1$.
Hence, when $\Lambda'$ and $\Lambda''$ are both integral \ie\ when
$c^2$ is a multiple of four, they are unimodular, being even or odd
according as $c^2/4$ is. This is a powerful result as odd unimodular
lattices can be found for all signatures, simply by considering
products of integers (hypercubic lattices). In fact this is the
simplest construction of an even unimodular lattice and it works
precisely when the signature $\eta$ is a multiple of eight since (4.8)
can be checked explicitly for the hypercubic lattices.

The lattice of free two-cycles $F_2(\M,\Z)$, endowed with the scalar
product furnished by the intersection number of pairs of two-cycles, was
unimodular. It was odd when $\M$ was Type II and even otherwise. In
the former case it must have a characteristic vector (4.7). This turns
out to be related to the quantity $w(\Sigma)$ introduced via the
quantum Stokes relation (3.3). Recall that 
$$
w(\Sigma)=I(\Sigma,\Sigma)
\hbox{ mod }2\quad\forall\,\Sigma\in F_2(\M,\Z).
\eqno(4.14)
$$
Now expand $\Sigma$ in terms of a basis $\Sigma=\sum_{j=1}^{b_2}n^j\Sigma_j$
and introduce the notations
$$
I(\Sigma_i,\Sigma_j)=(Q^{-1})_{ij}, \qquad w(\Sigma_i)=w_i,
\eqno(4.15)
$$
where the matrix $Q$ and its inverse both have integer entries as it
has determinant $\pm 1$. Now (4.14) reads
$$
\sum_{i,j,k=1}^{b_2}n^i(Q^{-1})_{ij}Q^{jk}w_k=
\sum_{i,j=1}^{b_2}n^i(Q^{-1})_{ij}n^j.
$$
This means that the characteristic vector of the lattice $F_2(\M,\Z)$
can be represented by the two-cycle
$$
\gamma=-\sum_{j,k=1}^{b_2}w_jQ^{jk}\Sigma_k.
\eqno(4.16)
$$
Actually this argument is the reverse of that used in our previous paper.

What will be more important is the  lattice structure of the quantised fluxes
through these two-cycles. A scalar product is provided by the theta term
contribution to the action (2.2). This is due to a generalisation
of the Riemann bilinear identity applicable to any pair of closed forms.
Here it reads
$$
\int_{\M}F\wedge F'=\sum_{i,j=1}^{b_2}\int_{\Sigma_i}F\, 
Q^{ij}\int_{\Sigma_j}F'.
\eqno(4.17)
$$
Considering first the quantised flux background necessary for $\M$ to support a
complex scalar wave function, it is natural to define field strengths
$F^1,F^2,\dots,F^{b_2}$ referring to a basis reciprocal to $\Sigma_i$:
$$
{q_B\over2\pi\hbar}\int_{\Sigma_i}F^j=\delta_i^j.
\eqno(4.18)
$$
The general solution to the quantisation condition (3.1) is
$$
F=\sum_{i=1}^{b_2}m_iF^i, \qquad m_i\in\Z
\eqno(4.19)
$$
(or something cohomologous). These de Rham cohomology classes form a
lattice with unimodular scalar product given by
$$
\left({q_B\over2\pi\hbar}\right)^2\int_{\M}F\wedge F
=\sum_{i,j=1}^{b_2}m_iQ^{ij}m_j.
\eqno(4.20)
$$
Thus the fluxes too form a unimodular lattice which is odd if $\M$
is of Type II and even otherwise.

Now consider the quantised flux backgrounds necessary for $\M$ to
support a complex spinor wave function and define a basis similar to
(4.18) but with $q_F$ replacing $q_B$. Then the general solution
to the quantisation condition (3.6) is
$$
F=\sum_{i=1}^{b_2}\left(m_i+{w_i\over2}\right)F^i, 
\eqno(4.21)
$$
(or something cohomologous), with scalar product
$$
\left({q_F\over2\pi\hbar}\right)^2\int_{\M}F\wedge F
=\sum_{i,j=1}^{b_2}\left(m_i+{w_i\over2}\right)Q^{ij}
\left(m_j+{w_j\over2}\right).
\eqno(4.22)
$$
It follows from (4.16) that $\sum_iw_iF^i$ represents the characteristic  
vector for  the unimodular flux lattice. Thus when $\M$ is of Type
I or III so that all $w_i$ vanish (mod $2$), the fluxes lie on
an even unimodular lattice. But when $\M$ is of Type II the fluxes
lie on an odd unimodular lattice displaced by half its characteristic vector,
that is the non trivial coset of (4.9).

In particular, if $q_B$ and $q_F$ are equal, the choice between fluxes
corresponding to the two terms in the decomposition (4.9) of $\Lambda_{TOT}$
depends on whether the complex wave function is scalar or
spinor. Notice that according to (4.7) one half of the expression
(4.22) differs from $c\cdot c/8$ by an integer. Then (4.8) would
follow from the integrality of the index of the Dirac operator on $\M$
using the version of the Atiyah-Singer index theorem quoted in [M
Alvarez and Olive]. 

\beginsection 5. The Maxwell partition function and  theta functions

We are now in a position to return to the evaluation of the partition
function (2.9), (2.10) in terms of the lattice structures just described.

The basic idea of the semi-classical approximation to (2.10) is to
expand the integrand about the stationary points of the exponent which
is given in terms of the classical action (2.2), that is, solutions to
the Euler-Lagrange equations, here the Maxwell equation 
${d*F=0}$. This together with $dF=0$ means that classical
solutions $F$ are harmonic two-forms on $\M$. Because the relevant
metric on $\M$ is Euclidean,  Hodge's theorem is valid and states that
there really is one and only one harmonic two-form, \ie\ stationary
point, in each cohomology class. As we saw these classes are labelled
by the $b_2$ magnetic integers $m_1,m_2,\dots, m_{b_2}$. It is
convenient now to suppose that the representative field strengths 
$F^1,F^2,\dots, F^{b_2}$ satisfying (4.18) are these harmonic
ones. Then the classical solutions in each class are given precisely
by (4.19) or (4.21), as the case may be. The space of harmonic
two-forms on $\M$ divides into a direct sum of two subspaces
consisting of self-dual and anti-self-dual harmonic two-forms. The
dimensions of these subspaces are the same numbers $b^+$ and $b^-$
previously defined in purely topological terms by means of the
$F_2(\M,\Z)$ intersection matrix $Q^{-1}$.  

The result is a sum over stationary points, that is the points of the
lattices  described in the previous section. The contribution of each
of these stationary points consists of the exponential of
${\im\over\hbar}$ times the Euclidean action $W_{EUCLIDEAN}$ evaluated
at the stationary point, all multiplied by a determinantal factor
$\Delta(\tau)$ formed by the Gaussian integral of the quadratic
fluctuations about it as well as zero modes. According to the
arguments of E Verlinde and Witten this determinantal factor is common
to all the terms of the sum. Furthermore, Witten argued that it takes the form 
$$
\Delta(\tau)=Z_0(\tau_2)^{{b_1-1\over2}}
\eqno(5.1)
$$
where $Z_0$ is independent of $\tau$ and $b_1$ denotes the first Betti number.
$\tau_1$ and $\tau_2$ are as defined in (2.3) but with $q$ replaced by
$\qb$ or $\qf$, as appropriate. Hence the partition function is given
by this factor $\Delta$ times a sum over a lattice of an exponential
of an expression quadratic in the coordinates of the lattice
points. This is recognisable as a sort of theta function associated
with the flux lattice which we shall examine in more detail. Because
the action (2.2) is quadratic in field strengths the result of the
procedure is exact and this is what makes this version of Maxwell
theory mathematically tractable.  

First we evaluate $\exp\left({\im\over\hbar}W_{EUCLIDEAN}\right)$ at
the stationary points (4.19) relevant to complex scalar wave functions
on any four manifold and to complex spinor wave functions on
four-manifolds of Types I and III. The contribution of the theta term
to the exponent is evaluated using (4.20): 
$$
{\im\tau_1\over2\hbar\tau_2}\int_{\M}F\wedge F=\im\pi\tau_1\,m^TQm.
\eqno(5.2)
$$
The evaluation of the Maxwell term relies on the fact that, if $F^i$
is harmonic, so is its dual $*F^i$. Hence there exists a matrix $G$ whereby
$$
*F^i=G^{ij}(Q^{-1})_{jk}F^k.
\eqno(5.3)
$$
Because of (2.1) and the Euclidean nature of the metric
$$
\left(GQ^{-1}\right)^2=1.
\eqno(5.4)
$$
Hence
$$
{1\over2\hbar}\int_{\M}F\wedge*F=-\pi\tau_2m^TGm. 
\eqno(5.5)
$$
In the course of this argument it becomes clear that the $b_2\times
b_2$ matrix $G$ is symmetric and positive definite, reflecting the
nature of the metric upon which it depends. Finally, evaluated at the
classical solution (4.19), 
$$
\e^{{i\over\hbar}W_{EUCLIDEAN}}=\e^{\im\pi m^T(\Omega(\hat\tau))m}.
\eqno(5.6)
$$
where
$$
\Omega(\tau)=\tau_1Q+i\tau_2G,
\eqno(5.7)
$$
Hence, for the backgrounds considered so far, the partition function
$$
Z(\tau)=\Delta(\tau)\Theta(\Omega(\tau)),
\eqno(5.8)
$$ 
where the richest structure resides in the theta function factor
which has the general form
$$
\Theta(\Omega)=\sum_{m_i\in\Z}\e^{\im\pi m^T\Omega m}.
\eqno(5.9)
$$
The sum converges if the complex symmetric matrix $\Omega$ has
imaginary part which is positive definite. This is certainly so in our case as
both $\tau_2$ and $G$ are positive in view of their physical interpretations. 

As a consequence of the Poisson summation formula, the theta function
(5.9) obeys the property 
$$
\Theta(-\Omega^{-1})=\sqrt{\det(-\im\Omega)}\,\,\Theta(\Omega),
\eqno(5.10a)
$$
where the positive sign of the root is understood. Furthermore
$$
\Theta(\Omega)=\Theta(A\Omega A^T)=\Theta(\Omega+B)
\eqno(5.10b)
$$
where $A\in GL(b_2,\Z)$ and $B$ is a symmetric matrix with integer
entries which are even on the diagonal. These symmetries generate a
subgroup of $Sp(2b_2,\Z)$ with finite index. We do not know whether
this has any physical significance but it does contain a subgroup recognisable 
as consisting of discrete electromagnetic duality transformations
acting on $\tau$ when we take account of some special properties
possessed by the matrix (5.7), namely 
$$
\Omega(-1/\tau)=-Q\Omega(\tau)^{-1}Q, \quad
\Omega(\tau+1)=\Omega(\tau)+Q,
\eqno(5.11a)
$$
and
$$
\sqrt{\det\left(-\im\Omega(\tau)\right)}=
e^{-{2\pi \im \eta\over8}}\tau^{b^+/2}(\tau^*)^{b^-/2},
\eqno(5.11b)
$$
checked using (5.4) and the fact that $GQ^{-1}$ has $b^{\pm}$ 
eigenvalues $\pm1$. As before, $\eta=b^+-b^-$ is the signature of $Q$
and hence the Hirzebruch signature of $\M$.

Regarding the theta functions as functions of $\tau$, and denoting
them accordingly as $\Theta(\tau)$, it follows from (5.10) and (5.11) that
$$
\Theta(-1/\tau)=e^{-{2\pi \im\eta\over8}}\tau^{b^+/2}(\tau^*)^{b^-/2}
\Theta(\tau)
\eqno(5.12a)
$$
$$\Theta(\tau+1)=\Theta(\tau) \qquad\hbox{if $Q$ is even},
\eqno(5.12b)
$$
and
$$\Theta(\tau+2)=\Theta(\tau) \qquad\hbox{otherwise},
\eqno(5.12c)
$$
We have already met the $S$-transformation (2.14). Together with 
$$
T=\pmatrix{1&1\cr 0&1\cr},
\eqno(5.13)
$$
sending $\tau\rightarrow\tau+1$, it generates the discrete group
$Sp(2,\Z)$, a subgroup of the $Sp(2,\R)$ duality group of section
2. On the other hand, $S$ and $T^2$ generate a subgroup of $Sp(2,\Z)$
called the Hecke group, $\Gamma_{\theta}$, one of the three distinct
subgroups of index three. 

So, roughly speaking, the partition function associated with
backgrounds of integrally quantised fluxes does manifest a property of
electromagnetic duality in that it transforms simply under certain
discrete subgroups of $Sp(2,\R)$ acting on the dimensionless variable
$\tau$. This discrete subgroup is $Sp(2,\Z)$ or $\Gamma_{\theta}$
depending on whether the lattice of integrally quantised fluxes is even or odd.
But it is precisely in the latter case that real spinor wave functions
are forbidden.

The existence of complex spinor wave functions requires fractionally
quantised fluxes, as explained above, and hence a different partition
function. Repetition of the above calculation using (4.21) rather than
(4.19) yields a similar result but with a different theta function, namely
$$
\Theta(\tau)_{w/2}=\sum_{m_i\in\Z}
\e^{\im\pi(m+{1\over2}w)^T\Omega(\tau)(m+{1\over2}w)},
\eqno(5.14)
$$
We shall show that this transforms nicely under the action of another
subgroup of $Sp(2,\Z)$, also of index three, like $\Gamma_{\theta}$.
Furthermore, if $\qf=\qb$, (5.14) is related to the previous theta
function, (5.9), by an element $ST$ of $Sp(2,\Z)$, outside the two
subgroups. Thus there is a sense in which the action of the full
modular group $PSL(2,\Z)\equiv Sp(2,\Z)/Z_2$ of electromagnetic
duality transformations can be realised taking into account the difference in
the background of quantised  fluxes needed to support complex scalar
and spinor wave functions respectively on four-manifolds of type II.

In order to explain this in more detail we need more developments in formalism.

Notice that although $\Theta(\Omega)$ is holomorphic in $\Omega$,
$\Theta(\tau)$ is not holomorphic in $\tau$ unless $b^-$ vanishes. For
then $G=Q$ and $\Omega(\tau)=\tau Q$. Similarly, if $b^+$ vanishes,
$\Omega(\tau)=\tau^*Q$. 

\beginsection 6. Theta functions and integral lattices

In this section the theta function construction above is extended in a way that
is similar to that described in the book [Green, Schwarz and Witten 1987]
but which goes further. It seems to be novel and intrinsically interesting. A theta function 
is associated to each element of the group $Z(\Lambda)$, (4.5), defined by
an integral lattice $\Lambda$, whatever the signature of its scalar product.
These $|Z(\Lambda)|$ theta functions support an action of the group
$Sp(2,\Z)$ (\ie\ $SL(2,\Z)$) if $\Lambda$ is even, and its Hecke
subgroup $\Gamma_{\theta}$ if it is odd (or, more properly, their
metaplectic extensions). 

Careful analysis of how the effect of the generators $S$ and $T$ of $Sp(2,\Z)$
satisfy the relation $(ST)^3=-I$ will yield \lq\lq Milgram's formula" 
expressing the signature (mod $8$) in terms of the structure of
$Z(\Lambda)$, when $\Lambda$ is even. The construction (4.11) means
that odd unimodular lattices can also be dealt with, leading to a
proof of (4.8). The relevance of all this to electromagnetic duality
will be explained in the following section.

We start with Poisson's summation formula in the form:
$$
\sum_{n_i\in\Z}f(x_i+n_i)= \sum_{m_i\in\Z}\e^{2\pi\im \,m_j x_j}
\tilde{f}(m_i),\quad\hbox{where}\quad 
\tilde{f}(k)= \int d^nx \,\e^{-2\pi\im \,k_j x_j}f(x)
\eqno(6.1)
$$
denotes the Fourier transform and the summation convention is understood.
The sums and integrals converge if
$$
f(x)=\e^{\pi \im\, x_j\Omega_{jm}x_m}\quad\hbox{so}
\quad\tilde{f}(k)={1\over\sqrt{\det(-\im\Omega)}}
\e^{-\pi \im\, k_j(\Omega^{-1})_{jm}k_m}
$$
and $\Omega$ is a complex symmetric matrix with positive definite imaginary
part, as before. Hence
$$
\sum_{n_i\in\Z}e^{\pi\im\,(x_j+n_j)\Omega_{jm}(x_m+n_m)}=
{1\over\sqrt{\det(-\im
\Omega)}}\sum_{m_i\in\Z}\e^{2\pi\im\,m_jx_j}
\e^{-\pi\im\,m_j\Omega_{jk}^{-1}m_k}.
$$
Now suppose that, as in (4.1), $n_j$ are the coordinates of the point
$l$ of the lattice $\Lambda$ with respect to the basis $e_i$, and
that $m_j$ are the coordinates of the point $l^*$ of the reciprocal
lattice $\Lambda^*$ with basis $f_j$. Thus 
$$
l=\sum_{j=1}^n n_je_j,\quad l^*=\sum_{j=1}^nm_jf_j\quad\hbox{while}\quad 
X=\sum_{j=1}^nx_je_j.
$$
Then, if we denote
$$
\hat\Omega=f_j\Omega_{jk}f_k^T, \quad\hbox{then}\quad
(\hat\Omega)^{-1}=e_j(\Omega^{-1})_{jk}e_k^T
$$
and the Poisson summation formula reads
$$
\sum_{l\in\Lambda}\e^{\pi\im\,(X+l)\cdot\hat\Omega\cdot(X+l)}=
{1\over\sqrt{\det(-\im\Omega)}}\sum_{l^*\in\Lambda^*}
\e^{2\pi\im\,l^*\cdot X}\e^{-\pi\im\,l^{*T}\cdot\hat\Omega^{-1}\cdot l^*}.
$$
Now suppose that the lattice $\Lambda$ is integral and that
the $Z(\Lambda)$ coset decomposition of $\Lambda^*$ can be written
$$
\Lambda^*=\Lambda\cup(\lambda_1+\Lambda)\cup(\lambda_2+\Lambda)\dots
\cup(\lambda_{|Z(\Lambda)|-1}+\Lambda), 
\eqno(6.2)
$$
where $\lambda_{\beta}$ is a representative element of the $\beta$'th
coset and $\lambda_0$ is understood to vanish. Choosing $X$ to be the
$\alpha$'th of these representatives so $\e^{2\pi\im\,l^*\cdot
X}=\e^{2\pi\im\,\lambda_{\alpha}\cdot\lambda_{\beta}}$ if
$l^*\in\lambda_{\beta}+\Lambda$, we can rearrange the Poisson
summation formula so that sums over $\Lambda$ occur on both sides:
$$
\sum_{l\in\lambda_{\alpha}+\Lambda}\e^{\pi\im\,l\cdot\hat\Omega\cdot l}=
{1\over\sqrt{\det(-\im\Omega)}}\sum_{\beta=0}^{|Z(\Lambda)|-1}
\e^{2\pi\im\,\lambda_{\alpha}\cdot\lambda_{\beta}}
\sum_{l\in\lambda_{\beta}+\Lambda}
\e^{-\pi\im\,l\cdot\hat\Omega^{-1}\cdot l}.
$$
Now consider $\Omega$ to have the special structure (5.7), where,
as before, $Q_{ij}=e_i\cdot e_j$, but is now only integral and not necessarily
unimodular. Then by the properties of the reciprocal basis
$$
\hat\Omega(\tau)=\tau_1I+i\tau_2\hat G\quad\hbox{where}
\quad\hat G=f_iG_{ij}f_j^T\quad\hbox{and}\quad \hat G^2=I
$$
by virtue of equations (5.4) and (2.1). It follows that
$\hat\Omega^{-1}(\tau)=\hat\Omega(1/\tau)$. Hence, defining the
$|Z(\Lambda)|$ theta functions 
$$
\Theta_{\alpha}(\tau)=\sum_{l\in\lambda_{\alpha}
+\Lambda}\e^{\pi\im\,l\cdot(\tau_1+\im\tau_2\hat G)\cdot l},
\quad \alpha=0,1\dots |Z(\Lambda)|-1,
\eqno(6.3)
$$
the Poisson summation formula now reads
$$
\Theta_{\alpha}(\tau)=\tau^{-b^+/2}(\tau^*)^{-b^-/2}{\e^{2\pi\im\eta/8}
\over\sqrt{|Z(\Lambda)|}}\sum_{\beta=0}^{|Z(\Lambda)|-1}
\e^{2\pi\im\,\lambda_{\alpha}\cdot\lambda_{\beta}}
\Theta_{\beta}(-1/\tau)
\eqno(6.4)
$$
where the determinant was evaluated by (5.11b), modified to take
account of the extra factor $|\det\, Q|=|Z(\Lambda)|$, by (2.10), arising
because $Q$ is no longer necessarily unimodular.

This is the action of the $S$-transformation (2.14). The response to
$T$, (5.13), is simple when $\Lambda$ is even. Otherwise $T^2$ must be
considered 
$$
\Theta_{\alpha}(\tau+1)=
\e^{\pi\im\,\lambda_{\alpha}^2}\Theta_{\alpha}(\tau)
\quad \hbox{if $\Lambda$ is even}
\eqno(6.5a)
$$
$$
\Theta_{\alpha}(\tau+2)=
\e^{2\pi\im\,\lambda_{\alpha}^2}\Theta_{\alpha}(\tau)
\quad \hbox{otherwise}
\eqno(6.5b)
$$
As mentioned earlier, the two matrices $S$, (2.14), and $T$, (5.13),
generate $Sp(2,\Z)\equiv SL(2,\Z)$, a discrete subgroup of the
original duality group, $Sp(2,\R)$, while $S$ and $T^2$ generate the Hecke
subgroup, $\Gamma_{\theta}$. $S$ and $T$ are not independent since they
satisfy the relations
$$
S^2=-I_2=(ST)^3.
\eqno(6.6)
$$
Given the responses (6.4) and (6.5) of the theta functions, the relations
(6.6) will have remarkable consequences that we now develop. First
define the following action of the general element $B=\pmatrix{a&b\cr
c&d\cr}$ of $Sp(2,\R)$ on a function $f(\tau)$ of $\tau_1$ and
$\tau_2$, (or equivalently $\tau$ and $\tau^*$),
$$
\hat A_{k_1k_2}(B)f(\tau)=(c\tau+d)^{-k_1}(c\tau^*+d)^{-k_2}
f\left({a\tau+b\over c\tau+d}\right).
\eqno(6.7)
$$
This is meaningful if $k_1$ and $k_2$ are are a pair of integers as
will be supposed. It is also relevant for $k_1$ and $k_2$ to be
half-integers. Treatment of this case requires the introduction of the
metaplectic group to take account of the sign ambiguities that arise
because of the square roots. In the interests of avoiding an over cumbersome 
notation this will not be done.

The virtue of the definition (6.7) is that this action satisfies the group
property:
$$
\hat A_{k_1k_2}(B)\hat A_{k_1k_2}(B')=\hat A_{k_1k_2}(BB').
\eqno(6.8)
$$
Given a column vector of such functions, $f_1(\tau),f_2(\tau)\dots f_N(\tau)$,
we say, following [V Kac 1990], that they form a vector modular form
if there exist matrices $D_{\beta\alpha}(B)$ such that
$$
\hat A_{k_1k_2}(B)f_{\alpha}(\tau)=
\sum_{\beta=1}^N f_{\beta}(\tau)D_{\beta\alpha}(B).
\eqno(6.9)
$$
The integers $k_1$ and $k_2$ are then called the weights of the functions
$f$ with respect to whatever discrete subgroup of $Sp(2,\Z)$ is considered.

It follows from (6.8) and (6.9) that, if the $N$ functions $f$ are
linearly independent, then the matrices $D$ represent the relevant group:
$$
D(B)D(B')=D(BB').
\eqno(6.10)
$$
Furthermore
$$
D_{\alpha\beta}(I_2)=\delta_{\alpha\beta}, \quad 
D_{\alpha\beta}(-I_2)=(-1)^{k_1-k_2}\delta_{\alpha\beta}.
\eqno(6.11)
$$
But the theta functions (6.3) are not necessarily linearly independent
as they may be related by permutation matrices (which are real). For example, 
$$
\Theta_{\alpha}(\tau)P_{\alpha\beta}=\Theta_{\beta}(\tau)\quad\hbox{where}
\quad P_{\alpha\beta}=\delta_{\lambda_{\alpha}+\lambda_{\beta},0},
\eqno(6.12)
$$
and it is understood in the Kronecker delta function that the equality to zero
is mod $\Lambda$. 

It follows from rearrangement of (6.4) and (6.5) that the $|Z(\Lambda)|$ 
theta functions (6.3) constitute a vector modular form with weights
$(k_1,k_2)=(b^+/2,b^-/2)$ and that, for them, 
$$
D_{\alpha\beta}(S)={\e^{-2\pi\im\eta/8}\over\sqrt{|Z(\Lambda)|}}
\e^{-2\pi\im\lambda_{\alpha}.\lambda_{\beta}},
\eqno(6.13)
$$
$$
D_{\alpha\beta}(T)=\e^{\pi\im\lambda_{\alpha}^2}\,\delta_{\alpha\beta}
\quad\hbox{if $\Lambda$ is even,}
\eqno(6.14a)
$$
$$
D_{\alpha\beta}(T^2)=\e^{2\pi\im\lambda_{\alpha}^2}\,\delta_{\alpha\beta}
\quad\hbox{otherwise.}
\eqno(6.14b)
$$
It is important to notice that in view of the assumed properties of
the lattice $\Lambda$, these matrix elements are independent of the
choice of representatives $\lambda_{\alpha}$ of the cosets
(6.2). Furthermore they are independent of the parameters in the
$b^+b^-$-dimensional moduli space of the matrices $G$.

The matrices (6.13) and (6.14) are unitary, obviously so for $D(T)$
and $D(T^2)$. The unitarity of $D(S)$  follows from the orthogonality 
properties of the characters of the abelian group $Z(\Lambda)$ which
appear as the phases
$\exp(2\pi\im\lambda_{\alpha}\cdot\lambda_{\beta})$. Actually the unitarity
of $D(S)$ was used in the derivation of (6.13) from (6.4).

By inspection,
$$
D(S)^T=D(S),\quad D^*(S)=(-1)^{\eta/2}PD(S)=(-1)^{\eta/2}D(S)P
\eqno(6.15)
$$
where $P$ is defined by (6.12). Hence
$$
D(S)^2=D(S)D(S)^{\dagger}(-1)^{\eta/2}P=(-1)^{\eta/2}P=D(-I_2)P
$$
using unitarity and (6.11). As $P$ equals unity on theta functions,
(6.12), the first of the relations (6.6) is therefore checked on the
theta functions (6.3). The other identity is more interesting. By
(6.4) and (6.5), 
$$
\left(\left(D(S)D(T)\right)^2\right)_{\alpha\beta}= 
{\e^{-4\pi\im \eta/8}\over |Z(\Lambda)|}
\,\sum_\gamma=0^{|Z(\Lambda)|-1}\e^{\pi\im(-2\lambda_\alpha\cdot\lambda_\gamma+
\lambda_\gamma^2-2\lambda_\gamma\cdot\lambda_\beta+\lambda_\beta^2)}.
$$
Rearranging the sum over $\gamma$, this can be written in the form
$$
\left(\left(D(S)D(T)\right)^2\right)_{\alpha\beta}
={\Psi(\Lambda)\e^{-2\pi\im \eta(\Lambda)/8} \over\sqrt{|Z(\Lambda)|}}
\,\e^{-\pi\im(2\lambda_\alpha\cdot\lambda_\beta+\lambda_\alpha^2)},
\eqno(6.16)
$$
where
$$
\Psi(\Lambda)\equiv {\textstyle{\e^{-2\pi\im\eta(\Lambda)/8}
\over\sqrt{|Z(\Lambda)|}}}
\sum_{\gamma=0}^{|Z(\Lambda)|-1}
\e^{\pi\im(\lambda_\gamma-\lambda_\alpha-\lambda_\beta)^2}=
{\textstyle{\e^{-2\pi\im\eta(\Lambda)/8}
\over\sqrt{|Z(\Lambda)|}}}\sum_{\gamma=0}^{|Z(\Lambda)|-1}
\e^{\pi\im(\lambda_\gamma)^2}.
\eqno(6.17)
$$
So the quantity $\Psi(\Lambda)$ is independent of $\alpha$ and $\beta$
by virtue of the way addition of cosets in (6.2) realises the group
$Z(\Lambda)$. Now we observe that
$$
\left(\left(D(S)D(T)\right)^{\dagger}\right)_{\alpha\beta}=
\left(D^*(S)D^*(T)\right)_{\beta\alpha}=
{\textstyle{1\over\sqrt{|Z(\Lambda)|}}}\e^{{1\over 8}2\pi\im \eta}\,
\e^{\pi\im(2\lambda_\alpha\cdot\lambda_\beta-\lambda_\alpha^2)}.
$$
This can be made proportional to $(D(S)D(T))^2$ by means of the matrix $P$,
which changes the sign of the vector $\lambda_\beta$:
$$
\left(\left(D(S)D(T)\right)^{\dagger}\,P\right)_{\alpha\beta}=
\left(D^*(S)D^*(T)P\right)_{\beta\alpha}=
{\textstyle{1\over\sqrt{|Z(\Lambda)|}}}\e^{{1\over 8}2\pi\im \eta}\,
\e^{-\pi\im(2\lambda_\alpha\cdot\lambda_\beta+\lambda_\alpha^2)}.
$$
Comparing the last equation with (6.16)  we find that
$$
\left(D(S)D(T)\right)^2=\e^{-4\pi\im \eta/8}\Psi\,
\left(D(S)D(T)\right)^{\dagger}\,P.
$$
The matrix $D(S)D(T)$ is unitary, and this implies that 
$$
\left(D(S)D(T)\right)^3=\e^{-4\pi\im \eta/8}\Psi\,P.
$$
But, by (6.11), $D(-I_2)=\exp(4\pi\im\eta/8)$ and therefore the result
we are looking for is
$$
D(-I_2)\left(D(S)D(T)\right)^3=\Psi(\Lambda)\,P.
$$
This has to be equal to the identity in order to satisfy the
relation $(ST)^3=-I_2$, (6.6). When acting on theta functions the
matrix $P$ is the identity, and this implies that
$\Psi(\Lambda)$, (6.17), must have a particular value, namely unity. Hence
$$
{\textstyle{1\over\sqrt{|Z(\Lambda)|}}}\sum_{\gamma=0}^{|Z(\Lambda)|-1}
\e^{\pi\im\lambda_\gamma^2}=
\e^{{1\over 8}2\pi\im \eta(\Lambda)}.
\eqno(6.18)
$$
This is Milgram's formula, relating the signature of any even lattice
to its structure, but so far only proven when $b^+$ and $b^-$ and
hence their difference, $\eta$, the signature of $\Lambda$, are all
even. But it is easy to check (6.18) explicitly for the even lattice
$\Lambda$ given by $\sqrt2\,\Z$, so $Z(\Lambda)=Z_2$, for either sign in
the natural scalar product. Furthermore, if $\Lambda_1\oplus\Lambda_2$
is the even lattice which is the orthogonal sum of the two even lattices 
$\Lambda_1$ and $\Lambda_2$ it is easy to check that
$\Psi(\Lambda_1\oplus\Lambda_2)=\Psi(\Lambda_1)\Psi(\Lambda_2)$. Thus
if two of the three lattices satisfies Milgram's formula so does the third.
Using this (6.18) can then be deduced for any even lattice.

In particular, Milgram's formula, (6.18), implies that the signature
of an even unimodular lattice is a multiple of eight as stated just
after equation (4.8). 

\beginsection 7. Theta functions and odd unimodular lattices

If the lattice $\Lambda$ is unimodular, the group $Z(\Lambda)$ consists
just of the unit element. Hence the previous construction (6.3) yields
just one theta function and the matrices $D$ are simply phase factors,
by unitarity, usually called multipliers. 

If, in addition, $\Lambda$ is even, the signature is a multiple of
eight, by Milgram's formula (6.18). Hence the multipliers (6.13) and
(6.14a) are trivial:
$$
D(S)=D(T)=1.
$$
If, instead, the unimodular $\Lambda$ is odd there is no restriction
on its signature. Again there is just one theta function supporting
transformations with respect to the Hecke subgroup, $\Gamma_{\theta}$,
generated by $S$ and $T^2$ but with non-trivial multipliers
$$
D(S)=\e^{-2\pi\im\eta/8} \qquad\hbox{ and }\qquad D(T^2)=1.
\eqno(7.1)
$$
However, associated with an odd unimodular lattice $\Lambda$ is the
even lattice $\Lambda_{EVEN}$ for which the group  $Z(\Lambda_{EVEN})$
consists of four elements, (4.11). The results of the previous section
suggest that there are three more theta functions which, altogether,
support an action under the full group $SL(2,\Z)$ which we now investigate.

It is this construction that will be precisely relevant to the Maxwell
partition functions associated with four-manifolds of Type II.

Since 
$$
SL(2,\Z)=
\Gamma_{\theta}\cup T\,\Gamma_{\theta}\cup ST\,\Gamma_{\theta},
\eqno(7.2)
$$
$\Gamma_{\theta}$ is a subgroup of index three in $SL(2,\Z)$ and so it
is natural to expect that only three theta functions are needed for
the complete action. Indeed this is true and the fourth theta function
can be chosen to be orthogonal to the three whilst supporting an
$SL(2,\Z)$ action on its own (if it does not vanish), as we shall see.

The following vectors are chosen as representatives of the cosets in the
decomposition (4.12) of $\Lambda_{TOTAL}=\Lambda_{EVEN}^*$:
$$
0,\qquad\lambda_v,\qquad 
\lambda_s=c/2,\qquad\lambda_t=\lambda_v+c/2,
\eqno(7.3)
$$
where $\lambda_v$ is simply any element of $\Lambda_{ODD}$. Then the
defining properties of the characteristic vector $c$ lead to the
conclusion that 
$$
{1\over2}\sum_{\gamma=0,v,s,t}\e^{\pi\im\,\lambda_{\gamma}^2}
={(1-1+2\e^{\pi\im\,c^2/4})\over2}=\e^{2\pi\im\,c^2/8}
$$
Using this, Milgram's formula (6.18) reduces to  (4.8) which is thereby proven.

The notation in (7.3) is the same as one customarily used for the
decomposition of the weight lattice of the Lie algebra
$so(2r)$. Corresponding to this $\Lambda$ is the hypercubic lattice $\Z^r$.
This is indeed a special case and many of the properties familiar in that case
will turn out to be true in much greater generality.

This is despite the fact that the theta functions associated with the odd 
unimodular Euclidean lattice $\Z^r$ take the form
$$
\left(\sum_{n\in\Z}\e^{\pi\im\,\tau n^2}\right)^r
$$
and no such factorisation occurs for more complicated odd unimodular lattices
such as $\Gamma_{8n+4}$ given by the construction (4.13), starting from the
hypercubic lattice $\Lambda=\Z^{8n+4}$, or the theta functions associated with
$\Z^r$ with indefinite scalar product.

Corresponding to the decomposition (7.3) there are four theta functions,
denoted, respectively $\Theta_0(\tau),\Theta_v(\tau),\Theta_s(\tau)$ 
and $\Theta_t(\tau)$. It is easy to evaluate the four-by-four matrices
$D(S)$ and $D(T)$ given by (6.13), (6.14a) and (7.3) in terms of 
$\omega=\e^{-2\pi\im\eta(\Lambda)/8}$ as
$$
D_{\alpha\beta}(S)={\omega\over2}
e^{-2\pi\im\lambda_{\alpha}\cdot\lambda_{\beta}}
={\omega\over2}\pmatrix{1&1&1&1\cr1&1&-1&-1\cr1&-1&\omega^2&-\omega^2\cr
1&-1&-\omega^2&\omega^2\cr}
\eqno(7.4)
$$
$$
D(T)=\hbox{diag}(1,-1,\omega^{-1},\omega^{-1}).
\eqno(7.5)
$$
It immediately follows from the definition (6.9) with (7.4) and (7.5) that
$$
\hat A(S)\left(\Theta_s-\Theta_t\right)=\omega^3\left(\Theta_s-\Theta_t\right)
\quad\hat A(T)\left(\Theta_s-\Theta_t\right)=\omega^{-1}\left(\Theta_s-\Theta_t
\right)
$$
Hence the linear combination $\Theta_s-\Theta_t$ supports an action of 
$SL(2,\Z)$ with the multipliers indicated. Quite often $\Theta_s-\Theta_t$
vanishes, for example if the signature $\eta(\Lambda)$ is odd or if
$\Lambda$ is Euclidean and hypercubic.

The sub-space orthogonal to $\Theta_s-\Theta_t$ is three dimensional
and spanned by $\Theta_0$, $\Theta_v$ and $\Theta_s+\Theta_t$. Of
course the theta function associated with the original odd unimodular
lattice $\Lambda$ is 
$$
\Theta=\Theta_0+\Theta_v
$$
and it is this which supports the action of the Hecke subgroup
$\Gamma_{\theta}$. It is convenient to define
$$
\Theta^T(\tau)\equiv\hat A(T)\Theta=\Theta_0(\tau)-\Theta_v(\tau)
$$
using (7.5), and
$$
\Theta^{ST}(\tau)\equiv\hat A(S)\Theta^T=\hat A(ST)\Theta=
\omega(\Theta_s(\tau)+\Theta_t(\tau)),
$$
using (7.4). Then $\Theta$, $\Theta^T$ and $\Theta^{ST}$ form an
alternative orthogonal basis for the three dimensional subspace in
which the action of $S$ is now represented by permutation matrices.

It is this basis that is relevant to the distinct Maxwell partition
functions arising when the space-time four-manifold is of Type II so that
the unimodular lattice $F_2(\M,\Z)$, (3.2), is unimodular and odd. The
partition function for backgrounds supporting complex scalar wave
functions is 
$$
Z_{SCALAR}(\tau)=\Delta(\tau)\Theta(\tau)
$$
while the partition function for backgrounds supporting complex spinor
wave functions is 
$$
Z_{SPINOR}(\tau)=\Delta(\tau)\Theta_{w/2}(\tau)=
\Delta(\tau)(\Theta_s(\tau)+\Theta_t(\tau))=
\omega^{-1}\Delta(\tau)\Theta^{ST}(\tau)
$$
using (5.14) and assuming that the respective electric charges $q_B$
and $q_F$ are equal. 

Of course $\Delta(\tau)\Theta^T(\tau)$ is simply $Z_{SCALAR}$ with
the angle $\theta$, (2.3), (2.4) increased by $2\pi$.

We have seen that the four dimensional space of theta functions
associated with the even lattice, $\Lambda_{EVEN}$ constructed from any odd 
unimodular lattice via (4.10) usually decomposes under the $SL(2,\Z)$
action as $4=3+1$, though sometimes, for example when the signature
$\eta(\Lambda)$ is odd, it is $4=3+0$ as $\Theta_s-\Theta_t$
vanishes. In fact when $\eta(\Lambda)$ is a multiple of four there is
a further decomposition $3=2+1$ whose details depend on whether 
$\eta(\Lambda)/4$ is odd or even. This is because of the occurrence of
the two new unimodular lattices (4.13) which are odd or even as 
$\eta(\Lambda)/4$ is. Associated with these are the theta functions 
$\Theta_0+\Theta_s$ and $\Theta_0+\Theta_t$ obeying
$$
\hat A(S)\left(\Theta_0+\Theta_s\right)=\omega(\Theta_0+\Theta_s),
$$
$$
\hat A(S)\left(\Theta_0+\Theta_t\right)=\omega(\Theta_0+\Theta_t),
$$
where, now, $\omega=(-1)^{\eta(\Lambda)/4}$. Also
$$
\hat A(T)(\Theta_0+\Theta_s)=\Theta_0+\omega\Theta_s,
$$
$$
\hat A(T)(\Theta_0+\Theta_t)=\Theta_0+\omega\Theta_t.
$$
So, when $\eta(\Lambda)$ is a multiple of eight, so $\omega=1$, the
three dimensional subspace contains the modular function 
$$
2\Theta_0+\Theta_s+\Theta_t=\Theta+\Theta^T+\Theta^{ST},
$$
with multipliers $D(S)=1=D(T)$. On the other hand when
$\eta(\Lambda)\in8\Z+4$, so $\omega=-1$ it is 
$$
2\Theta_v-\Theta_s-\Theta_t=\Theta-\Theta^T+\Theta^{ST}
$$
which has multipliers $D(S)=D(T)=-1$. 

\beginsection 8. Discussion

We have seen that the response of the extended partition function $Z$ 
associated with a four-manifold $\M$ depends on the topological
properties of $\M$. For example, if $\M$ is of type I or III
(according to the terminology of section 3), so that $F_2(\M,\Z)$ is
an even unimodular lattice, there is just one partition function and
it transforms under the full electromagnetic duality group $SL(2,\Z)$
acting on the dimensionless electromagnetic coupling $\tau$, (2.4), by
fractional linear transformations. If $\M$ is of type II so that
$F_2(\M,\Z)$ is an odd unimodular lattice there are three possible
partition functions, two appropriate to fluxes supporting complex
scalar wave functions and the third, complex spinor wave functions. 
Individually these transform under three distinct but conjugate
subgroups of $SL(2,\Z)$ of index three (one of which is the Hecke subgroup).
The full electromagnetic duality $SL(2,\Z)$ is realised by
permutations of these three, as explained above. 

In talking of transformations we allow for the possiblity of non-zero 
\lq\lq modular weights". We saw after equation (6.12) that the modular
weights of the theta functions were given by $b^{\pm}(\M)/2$. Witten [95]
argued that the prefactors due to the Van Vleck determinants (5.1) had
modular weights both given by $(1-b_1(\M))/2$. Hence the total modular
weights of $Z$ are
$$
{1-b_1(\M)+b^{\pm}(\M)\over2}={\chi(\M)\pm\eta(\M)\over4},
\eqno(8.1)
$$
using (2.13) and (4.2). These numbers possess several interesting properties.
First they are integers (or maybe half-integers in which case we
should properly talk of metaplectic versions of $SL(2,\Z)$). Secondly
they are rather special topological numbers possessing the property of 
\lq\lq locality" in the sense that the can be expressed as integrals over $\M$
of local densities.

Suppose space-time $\M$ exhibits a discrete $Z_2$ symmetry with no
fixed points. Then the quotient $\M/Z_2$ obtained by identifying
related points is also a four-manifold and it follows from the
locality properties that 
$$
\chi\left(\M/Z_2\right)={\chi(\M)\over 2}.
\eqno(8.2)
$$
If the quotient $\M/Z_2$ is Poincar\'e dual, the same relation (8.2)
holds for the Hirzebruch signature $\eta$ and hence for the modular
weights (8.1). 

If the partition function considered is a \lq\lq strict" one, that is
it is indeed a trace, (2.9), so that $\M$ has the form $S^1\times{\cal
M}_3$, then, as $S^1\equiv S^1/Z_2$, $\M\equiv\M/Z_2$, where the $Z_2$
relates diametrically opposite points on the circle. Consequently, by
(8.2), $\chi$, $\eta$ and the modular weights (8.1) all vanish and the
partition function is actually invariant, in agreement with initial
expectation. It is easy to check by calculation that
$F_2(S^1\times{\cal M}_3,\Z)$ is an even lattice so that type II is ruled out.
 
In general, the extended partition functions described in section 2 as
being associated with more general four-manifolds are only modular
covariant as the weights (8.1) need not vanish. This result is clearly
of interest, but it has to be admitted that it is unclear what its
physical meaning is. One reason is that a choice has to be made in
selecting the Euclidean metric on $\M$. Fortunately the dependence on
that choice is not too strong. For example, if the metric is altered
by a Weyl rescaling, 
$$
g_{\mu\nu}(x)\rightarrow\lambda(x)^2g_{\mu\nu}(x), 
\eqno(8.3)
$$
then $*F$ is unaltered if $F$ itself is.  Consequently the matrix $G$
defined by (5.3) is unchanged as is the complete theta
function. Perhaps this is related to the fact that when $b^+$ or
$b^-$ vanishes, then $G=\pm Q$ and so all dependence on the metric
disappears from the theta function (which is then holomorphic or
anti-holomorphic in $\tau$). If $b^+ $ and $b^-$ both vanish the
action (2.1) is independent of $\theta$, as mentioned earlier. Now we
see that the theta function is then simply a constant independent of $\tau$.

Of course, if $\M$ supports a Minkowski metric, its Euler number, $\chi$,
vanishes and the modular weights (8.1) become equal and opposite.

We should like to emphasise that the breakdown of $SL(2,\R)$ to its
discrete subgroup $SL(2,\Z)$ was a consequence simply of the Dirac
quantisation condition without recourse to the Zwanziger-Schwinger
condition, which, accordingly, seems less fundamental. 

In this paper we have endeavoured to elucidate the conceptual and
mathematical structure in what seems to be the simplest possible
context for electromagnetic duality. It would be physically
interesting to extend the work in many different ways, for example: 
\smallskip
1) to consider a larger number of abelian gauge fields on $\M$.

2) to consider space-times ${\cal M}_{4k}$ with $2k$-form field strengths
with potential couplings to $(2k-2)$-branes, possibly spinning.

3) to consider space-times ${\cal M}_{4k+2}$ with several $2k+1$-form
field strengths.  

4) to consider open space-times with boundary.

5) to consider the effect of space-time topology on conventional
superstring theories.  
\smallskip
Undoubtedly yet more mathematical tools would have to brought into play.

\bigskip

\centerline{\bf Acknowledgements}
\medskip
DIO wishes to thank Gary Gibbons, Stephen Howes, Nadim Mahassen, Boris
Pioline for discussions. He also wishes to thank the Mittag-Leffler
Institute, UNESP Institute for Theoretical Physics (S\~ao Paulo) and
the Instituut voor Theoretische Fysica, Utrecht
for hospitality and their members for congenial and enlightening
discussions. MA's research has been supported by EPSRC and
subsequently by PPARC through its Special Programme Grant
PPA/G/S/1998/00613. We both wish to thank TMR grant FMRX-CT96-0012 for
assistance.  
\medskip
\parindent=0pt
{\bf References} 
\medskip
M Alvarez and DI Olive; \lq \lq The Dirac quantisation
condition for fluxes on Four-manifolds", {\tt hep-th/9906093}, to
appear in \cmp\
\smallskip
O Alvarez; \lq\lq Topological Quantisation and Cohomology",
\cmp\ {\bf 100} (1985) 279-309 
\smallskip
E Cremmer and B Julia; \lq\lq The $SO(8)$
Supergravity", \np\ {\bf B159}, (1979) 141-212
\smallskip
A D'Adda, R Horsley and P Di Vecchia; \lq\lq Supersymmetric Monopoles
and Dyons", \hfil\break\pl\ {\bf B76} (1978) 298-302.
\smallskip
D Bohm and Y Aharonov; ``Significance of electromagnetic
potentials in the quantum theory'', \pr\ {\bf{115}} (1959), 485
\smallskip
PAM Dirac; \lq\lq Quantised singularities in the
electromagnetic field", \prs\ {\bf A33} (1931) 60-72
\smallskip
PAM Dirac;
``Gauge invariant formulation of Quantum Electrodynamics'', {\it Canadian
Journal of Physics} {\bf{33}} (1955), 650-660
\smallskip
RP Feynman and AR Hibbs; {\sl Quantum Mechanics and Path
Integrals}, Chapter 10, \hfil\break McGraw-Hill, 1965
\smallskip
RP Feynman; {\sl Statistical Mechanics}, Chapter 3,
Benjamin/Cummings, 1972
\smallskip
MK Gaillard and B Zumino; \lq\lq Duality Rotations for 
Interacting Fields", \np\ {\bf B193} (1981), 221-244 
\smallskip
M Green, J Schwarz and E Witten; {\sl Superstring Theory},
Vol.~ 2, Appendix 9B, Cambridge University Press, 1987
\smallskip
M Kalb and P Ramond; \lq\lq Classical direct interstring action",
\pr\ {\bf D9} (1974), 2273-2284
\smallskip
V Kac; {\sl Infinite dimensional Lie algebras}, Chapter 13,
Cambridge University Press, 1990
\smallskip
HA Kramers and GH Wannier; \lq\lq Statistics of the Two-Dimensional
Ferromagnet I",\hfil\break \pr {\bf 60} (1941), 252-276
\smallskip
HB Lawson and M-L Michelsohn; {\sl Spin Geometry},
{\it Princeton Mathematical Series} {\bf 38}, Princeton, 1989
\smallskip
J Milnor, D.~Husemoller; {\sl Symmetric bilinear 
forms}, Ergebnisse der Mathematik und ihrer Grenzgebiete, Band 73, 
Springer-Verlag, 1973.
\smallskip
C Montonen and DI Olive; \lq\lq Magnetic monopoles as gauge
 fields?", \pl\ {\bf B72} (1977), 117-120.
\smallskip
H Osborn; ``Topological Charges for $N\!=\!4$ Supersymmetric
Gauge Theories and Monopoles of Spin 1", \pl\ {\bf B83} (1979), 321-326. 
\smallskip
A Schwarz; {\sl Topology for Physicists}, Springer, 1994.
\smallskip
A Sen; \lq\lq Dyon-monopole bound states,
self-dual harmonic forms on the multi-monopole moduli space, and
$SL(2,Z\!\!\!\!Z)$ invariance in string theory", \pl\ {\bf B329}
(1994), 217-221 
\smallskip
E Verlinde; \lq\lq Global aspects of electric-magnetic duality",
\np\ {\bf B455} (1995), 211-228.
\smallskip
K Wilson; ``Confinement of quarks'', \pr\ {\bf{D10}} (1974), 2445-2459
\smallskip
E Witten and DI Olive; \lq\lq Supersymmetry Algebras 
that Include Topological Charges", \pl\ {\bf B78} (1978), 97-101.
\smallskip
E Witten; \lq\lq On S-duality in abelian gauge theory",
{\it Selecta Math (NS)} {\bf 1} (1995), 383-410, {\tt hep-th/9505186}
\smallskip
TT Wu and CN Yang; \lq\lq Concept of non-integrable phase factors
and global formulation of gauge fields", \pr\ {\bf D12}
(1975), 3845-3857.

\bye